\documentclass[12pt]{article}
\usepackage{amssymb,graphicx}
\usepackage{amsmath}
\usepackage{epsfig}

\def\e{{\rm e}}

\def\l{\left(}
\def\r{\right)}

\newcommand{\be}{\begin{equation}}
\newcommand{\ee}{\end{equation}}
\newcommand{\bea}{\begin{eqnarray}}
\newcommand{\eea}{\end{eqnarray}}
\newcommand{\bg}{\begin{gather}}
\newcommand{\eg}{\end{gather}}
\newcommand{\bseq}{\begin{subequations}}
\newcommand{\eseq}{\end{subequations}}

\renewcommand{\ln}{\mathop{\rm ln}\nolimits}

\begin{document}
\begin{center}
  {\Large\bf Star tracks in the ghost condensate} \\
\medskip
S.L.~Dubovsky$^{a,b}$\\
\medskip
  $^a${\small
Department of Physics, CERN Theory Division, CH-1211 Geneva 23, Switzerland
  }
\medskip
\\
$^b${\small
Institute for Nuclear Research of
         the Russian Academy of Sciences,\\  60th October Anniversary
  Prospect, 7a, 117312 Moscow, Russia
  }

\end{center}
\begin{abstract}
We consider the infrared modification
of gravity by  ghost condensate. Naively, in this scenario one expects
sizeable modification of gravity at distances of order 1000 km,
provided that  the
characteristic time scale of the theory is of the
order of the Hubble time.
However, we argue that this is not the case.
The main physical reason for the conspiracy is a simple
fact that the
Earth (and any other object in the Universe) has  velocity of at least of
order $10^{-3}c$ with 
respect to the rest frame of  ghost condensate.  
Combined with strong retardation effects present in the ghost sector, 
this fact
implies that no observable modification of the gravitational field of 
nearby objects occurs. Instead, the physical manifestation of ghost
condensate is the presence of ``star tracks'' --- narrow regions of space
with growing gravitational and ghost fields inside ---
 along the trajectory of
any massive object. We briefly discuss the possibilities
 to observe these tracks.
\end{abstract}
\section{Introduction}
\label{intro}
Emerging evidence for the accelerated expansion of the Universe triggered
interest in the non-standard theories of gravity in which gravitational
interactions get modified in the infrared. To some extent the motivation is
that in these theories, unlike in the case of pure cosmological constant, 
physics responsible for the cosmic acceleration may manifest itself in
observations at smaller distance scales (e.g., in Lunar Ranging
experiments \cite{Lue:2002sw,Dvali:2002vf}).

Naively, the most straightforward way to modify gravity at distance scale
$r_c$ 
would be
 to give a graviton a mass $m_g\sim r_c^{-1}$. However, conventional
massive gravity suffers either from the presence of ghosts or from the loss of
predictivity because of 
strong coupling at  unacceptably low energy scales
\cite{Arkani-Hamed:2002sp}. Similar problems arise in  multi-dimensional
models, and it is not clear whether there exists a  
consistent {\it quantum}
brane world theory where gravity is modified in the infrared and
predictive power is not lost at unacceptably low energy scale.
 
Recently, an
example of a theory which does not suffer from the
above problems has been constructed \cite{Arkani-Hamed:2003uy}. 
This theory, dubbed ``ghost condensate'', is somewhat
 similar to the Fierz--Pauli
massive gravity. The difference is that  
 the Fierz--Pauli mass term breaks
 reparametrization invariance completely, while
in the ghost condensate theory only the time reparametrization invariance
\[
t\to t+\xi(t,x)
\]
is broken, while the invariance under  
(possibly time-dependent) spatial diffeomorphisms is kept intact. 

As a result,  the latter 
theory becomes
formally reparametrization invariant after a single 
St\"uckelberg 
field is introduced, as opposed to four St\"uckelberg fields in
the Fierz--Pauli gravity. The key difference between these two theories is
that in the case of ghost condensate,  decoupling limit exists in
which 
gravity is switched off while the St\"uckelberg sector is still described by a
well defined low-energy effective theory valid up to a certain 
energy scale $M$.

The price  one pays is that the Lorentz invariance 
is not preserved by ghost condensate; 
for instance, at the quadratic level one effectively adds
to the Einstein theory the ``mass term'' of the form
\be
\label{mass}
\int dtd^3x{1\over 8}M^4h_{00}^2\;.
\ee
As a consequence of this violation of the Lorentz invariance,
 the dispersion law for the 
St\"uckelberg field\footnote{In what follows, we somewhat loosely refer to this
  field as a ghost. It is worth stressing however, that this is not a ghost
  field in the usual sense, i.e. the sign in front of its kinetic term in the
  action is positive.} $\pi$ has rather peculiar form,
\[
\omega^2={\alpha\over M^2} k^4\;,
\]
where $\alpha$ is a dimensionless parameter of the theory.
Another manifestation of the fact that the
 Lorentz invariance is broken is that
modification of gravity in the infrared is not characterized by a single
length scale $r_c$. Instead, there is a length scale $r_c$, equal to
\be
\label{rc}
r_c={\sqrt{2}M_{Pl}\over M^2}
\ee
and a time scale $t_c$ given by
\be
\label{tc}
t_c={2M_{Pl}^2\over \alpha M^3}\;.  
\ee
As discussed in
Ref.~\cite{Arkani-Hamed:2003uy}, for a massive source {\it at rest}
the
length scale
 $r_c$
determines the characteristic distance at which the 
gravitational potential
starts to deviate from the Newtonian one, while $t_c$ determines the
characteristic time needed for this deviation to show up. Naively, this
implies that, assuming that $\alpha\sim 1$ and $t_c$  is of the order
of the
present age of the Universe, $t_U\sim 15$ Gyr, one might expect a sizeable
modification of gravity due to ghost condensate at length scales of order 1000
km.

However,  it was noted already in Ref.~\cite{Arkani-Hamed:2003uy} that the
retardation effects
are very strong in the ghost sector. The point  
is that it is the whole
history of a system that determines its actual
gravitational potential in the presence of  ghost condensate. 
The purpose of this paper is to better understand this feature
and thus reconsider possible observational
signatures of ghost condensate.
It is worth noting that 
for the moment,
consequences of  ghost condensate with $t_c \sim t_U$ for
present day cosmology have not been elaborated yet
(see, however, 
Ref.~\cite{Arkani-Hamed:2003uz} where ghost condensate was used to 
construct a model of inflation
with quite unusual perturbation spectra). Still, we believe that the question
we address is of relevance, since  
ghost condensate is an interesting
infrared modification of gravity whose consistency is beyond any doubt.

Surprisingly, we find that the above naive expectation is incorrect
and it 
is not excluded 
that we  live in the Universe with $t_c\sim t_U$ and $r_c\sim
1000$~km and have not noticed that 
so far. To understand how that could be, one
recalls a simple fact
that objects in the Universe
are not at rest. Instead, solar system
and other stars in our Galaxy rotate around the center of the Galaxy with 
typical velocity of order $10^{-3}$, while the
Galaxy itself moves in the local
cluster of galaxies with the velocity of the same order of  magnitude.
 A
well-known observational consequence of this motion is the dipole anisotropy
of the cosmic microwave background.

This implies that
all stellar objects have velocity of at least of the same order of
magnitude with respect to the rest frame of  ghost condensate\footnote{A
  priori one may think that this velocity can be significantly 
larger if the rest frame
  of the CMB has a finite velocity with respect to the rest frame of the ghost
  condensate. On the other hand we expect that one of the effects
of the Hubble friction is to slow down this overall motion of the CMB. 
Throughout this paper we
assume that the rest frame of the CMB coincides with the rest frame of the
ghost condensate, leaving the study of whether it may be not
the case for future.}.
Now, it takes time of order $t_c$ for the modification of the gravitational 
potential to occur. Consequently, the effect of 
ghost condensate which we can observe on the Earth now is not a modification,
say, of the gravitational field of the Sun, but the ``ghost'' tail of the
potential of a star which was located nearby (in the rest frame of  ghost
condensate) time $t_c$ ago. The Universe with  
ghost condensate can be thought of as a kind of
a bubble chamber
where all moving massive objects leave  long (and, as we will see later,
narrow) tracks in which ghost field and gravitational potential are perturbed.
The time delay between the moment when the object passes  a given point
in space
and the appearance of the track around this point is of order $t_c$.
 
This observation is our main result. The rest of this paper is organized as
follows. In section \ref{general} we derive  a general expression for the 
gravitational potential of a massive source in the presence of ghost
condensate (in the Newtonian approximation). In section \ref{specific} we
consider specific examples, namely a potential of a moving
point-like source and the effect of the finite size of the source. 
Section \ref{evidence} contains preliminary discussion
of  
phenomenological implications of our calculations. 
Technical details can be found in  Appendix.
\section{Gravitational potential in 
the presence of ghost condensate}
\label{general}
Throughout this paper we consider 
small metric perturbations near flat
Minkowski space-time
\[
g_{\mu\nu}=\eta_{\mu\nu}+h_{\mu\nu}
\]
where $\eta_{\mu\nu}$ is the Minkowski metric with the signature $(+,-,-,-)$.
Excitations of  ghost condensate are described by a real scalar field
$\pi$, so, in the linear regime,
the modification of gravity takes place only in the scalar sector.
In the conformal Newtonian gauge the scalar part of the metric perturbation
$h_{\mu\nu}$  has the following non-zero components 
\[
h_{00}=2\Phi\;,\;\; h_{ij}=2\Psi\delta_{ij}\;.
\]
The quadratic Lagrangian describing the coupled system of scalar metric and
ghost condensate excitations is
\[
{\cal L}={\cal L}_{EH}+{\cal L}_{gh}+{\cal L}_{s}\;,
\]
where ${\cal L}_{EH}$ is the quadratic Einstein action in the Newtonian gauge,
 ${\cal L}_{gh}$ is the quadratic action for ghost condensate and ${\cal
L}_{s}$ is the source term.  Switching to the momentum space and taking the
Newtonian limit $\omega^2\ll k^2$, one has $\Phi=\Psi$ and the resulting
Lagrangian in the rest frame of  ghost condensate takes the following
form~\cite{Arkani-Hamed:2003uy} 
\begin{equation}
\label{action}
{\cal L}={1\over 2}\l \pi_c\;\Phi_c\r
\l
\begin{array}{cc}
\omega^2-\alpha^2k^4/M^2 & -im\omega\\
im\omega & -k^2+m^2
\end{array}
\r
\l
\begin{array}{c}
\pi_c\\
\Phi_c
\end{array}
\r+{\cal L}_{s}
\end{equation}
where the 
canonically normalized gravitational potential $\Phi_c$ is related to the
conventional one in the following way
\[
\Phi_c=\sqrt{2}M_{Pl}\Phi\;
\]
In Eq.~(\ref{action}) we use the notation
\[
m\equiv r_c^{-1}={M^2\over \sqrt{2}M_{Pl}}
\]
In what follows we consider sources with the energy-momentum of
the form
\[
T_{\mu\nu}=\rho u_\mu u_{\nu}
\]
In the non-relativistic limit when 
 the source velocity with respect to the rest frame of 
ghost condensate is small, one takes
\[
T_{00}=\rho(x,t)\;
\]
and sets all other components of $T_{\mu\nu}$ 
equal to zero. As a result one arrives
at the following source term in the action
\begin{equation}
\label{sourceterm}
{\cal L}_s=-{\sqrt{2}\over M_{Pl}}\Phi_c\rho
\end{equation}

 After 
inverting 
the $(2\times 2)$ matrix in Eq.~(\ref{action}) one finds that the
gravitational potential 
of a source can be written as a sum of two terms
\be
\label{NDel}
\Phi=\Phi_{N}+\Delta\Phi
\ee
where $\Phi_{N}$ is the conventional Newtonian potential and 
\begin{equation}
\label{generalphi}
\Delta\Phi(x,t)=-{2 G\over \pi t_c}\int^t
dt'\int d^3x'{\rho(t',x')\over |x-x'|}I\l{t-t'\over t_c},m|x-x'|\r
\end{equation}
The ``propagator'' $I(T,R)$ is
\begin{equation}
\label{ITR} 
I(T,R)=
\int_0^\infty {du\over\sqrt{u^2-1}}\sin{\l Tu\sqrt{u^2-1}\r}\sin{R u}
\end{equation}
In  Appendix we calculate various asymptotics of this function. The 
results are summarized as follows
\bseq
\label{ITRasympt}
\begin{align}
I(T,R)&={1\over 2}\sqrt{\pi\over T}\exp\l{T\over 2}-
{R^2\over 8T}\r\sin{R\over\sqrt{2}}\;,\;\; R\ll T\;,\;\; T\gg 1
\label{ITR1}
\\
I(T,R)& 
=\sqrt{\pi T\over 2}{\cos\l{R^2\over 4T}+{T\over 2}\r+\sin\l{R^2\over 4T}
+{T\over 2}\r\over R}\;,\;\;
R\gg T\;,\;\; R\gg 1
\label{ITR2}
\\
I(T,R)&\propto TR\;,\;\;R,T\ll1 
\end{align}
\eseq
It is straightforward to check numerically that in the 
intermediate regions the
function $I(R,T)$ smoothly interpolates
between the different asymptotics.
\section{Gravitational potentials of moving sources}
\label{specific}
\subsection{Point-like mass}
Now we are ready to calculate the
 gravitational potentials induced by different moving matter
sources. 
To begin with, let us find the potential of a point-like mass $\mu$ 
moving with  velocity $v \ll 1 $
with respect to ghost condensate
along the $z$ axis. The corresponding
mass density is equal to
\[
\rho(x,t)=\mu\delta^2(y)\delta(z-vt)
\]
where $y_{1,2}$ are transverse coordinates.
Plugging this source into Eq. (\ref{generalphi}) and integrating out
$\delta$-functions one gets
\begin{equation}
\label{generalpoint}
\Delta\Phi_\delta (y,z,t)
=-{2G\mu\over\pi vt_c}\int_0^{vt}{dz'\over
  r(z')}I\l{vt-z'\over vt_c},{r(z')\over r_c}\r
\end{equation} 
where 
\[
r(z')=\sqrt{y^2+(z'-z)^2}
\]
and 
\[
y^2=y_1^2+y_2^2
\]
is the distance from the observer to the trajectory of the source.
The potential (\ref{generalpoint}) 
crucially depends on whether the distance to the source trajectory $y$
 measured in  units of the characteristic length scale $r_c$
is large or small compared to the time interval (measured in units of $t_c$)
\[
T(t,z)={vt-z\over vt_c}
\]
between the moment of observation and the moment when the source was
close to
the point of observation\footnote{In what follows we assume that
$0<z<vt$, i.e. that the source has really passed the nearest point to
the observer. It is
straightforward to work out the expressions for the potential in other 
cases,
which are less interesting.}.  For 
\be
\label{limit1}
y\gg r_cT
\ee
 one makes use of the
asymptotics Eq.~(\ref{ITR2}). The integral in Eq.~(\ref{generalpoint}) is
saturated in the interval of width $\sim r_c\sqrt{T}$ near the
point $z'=z$, and one arrives at the following expression 
for the
potential in this case (barring
an extremely slowly varying $y$-independent phase of oscillations)
\begin{equation}
\label{weaklimit}
\Delta\Phi_\delta= -{2\sqrt{2}G\mu Tr_c^2\over y^2vt_c}
\sin{y^2
\over
  4r_c^2T}
\end{equation}
If, on the other hand, 
\be
\label{limit2}
y\ll r_cT
\ee
and $T\gg 1$, then using 
short distance asymptotics   
(\ref{ITR1}) of the propagator one obtains for the potential
\begin{equation}
\label{largepoint}
\Delta\Phi_\delta=-{G\mu\over\sqrt{\pi} vt_c}\int_0^{vt}{dz'\over
  r(z')\sqrt{T(t,z')}}\sin{r(z')\over
  \sqrt{2}r_c}\e^{{T(t,z')/ 2}}
\end{equation} 
This integral is also saturated in  the small interval near $z'=z$,
\[
|z-z'|\ll r_c T
\]
Then, to further simplify  the integral in Eq.~(\ref{largepoint}), 
one notes that for reasonable values of the parameters one has
\be
\label{naturallimit}
{vt_c\over r_c}={2vM_{Pl}\over \alpha M}\gg 1. 
\ee
Consequently, one can substitute $T(t,z')$ by $T(t,z)$ in
Eq.~(\ref{largepoint}). In the resulting integral one can extend 
the limits of
integration to $z'\in(-\infty,\infty)$ and  in this way one arrives
at the
following expression for the potential
%
\begin{equation}
\label{largepoint2}
\Delta\Phi_\delta=-\sqrt{\pi\over T}
{G\mu\over2 vt_c}
\e^{{T/2}}J_0\l{y\over \sqrt{2}r_c}\r
\end{equation}
This result is somewhat different from one obtained in
Ref.~\cite{Arkani-Hamed:2003uy}. Actually, it is easy to see that
 there is no contradiction. To
obtain the latter result from our Eq.~(\ref{largepoint})
consider the limit of extremely low velocity, such that 
\[
vt\ll y\;.
\]
In particular, this implies that
 a condition
opposite to our Eq.~(\ref{naturallimit}) holds. Then the interval of
 integration in Eq.~(\ref{largepoint}) is small and one can set
$r(z')=r(z)$. The resulting integral gives
\[
\Delta\Phi_\delta\propto G{\mu\over \sqrt{T}r}e^{T/2}\sin{r\over\sqrt{2}r_c}
\]
 This result is in agreement with
Ref.~\cite{Arkani-Hamed:2003uy}, as it should be. 

To summarize our results, the gravitational 
potential of the moving point-like mass in the
linear approximation can be
described as follows. There is a narrow cone behind the source 
(``track''), of  angle  
\[
\alpha=r_c/(vt_c)
\]
 where the potential oscillates in
the transverse direction and exponentially grows backward 
along the axis of the cone. The period of oscillation is equal to
$2\sqrt{2}\pi r_c$. The amplitude of oscillations decreases as $1/\sqrt{y}$
inside the cone, as the observer moves away from its axis. 
Outside the track there is a wave zone, where  nearly
cylindrical outgoing
wave of the gravitational potential is present. As one observes by
 expanding
 the argument of  sine in
Eq.~(\ref{weaklimit}) near a given space-time point, 
the frequency $\omega$ and the 
inverse wavelength $\lambda^{-1}$ of the wave grow with the distance from
the trajectory of the source,
\be
\label{wave}
\omega\sim{y^2\over r_c^2T^2t_c}\;,\;\; \lambda^{-1}
\sim{y\over
 r_c^2T} 
\ee 
The amplitude of this wave decreases as $1/y^2$. At small times $t<t_c$ the
exponential track is absent, and there is  the outcoming cylindrical
wave only.
\subsection{Source of  finite size}
\label{thicksub}
To discuss the observational
consequences  of  ghost condensate, one should extend the 
previous analysis to  sources of finite size.
Indeed, the
potentials (\ref{weaklimit}), (\ref{largepoint2}) rapidly oscillate in
space, so one expects that, unlike in  conventional gravity,
the amplitude of the potential in the track is not just proportional to the
mass of the source. This amplitude is expected to
be smaller for large sources due to
the effect of averaging.
  For simplicity we consider cylindrically symmetric sources described by the
following mass density,
 \be
\label{cyldens}
\rho_c={\mu\over L^2H} f(y/L)g((z-vt)/H)\;,
\ee
where the
functions $f$ and $g$ characterize the transverse and longitudinal 
profiles of the source,
 with $L$ and $H$ being the corresponding characteristic
length scales. 
$\mu$ is the total mass of the source, so that the following
normalization conditions are assumed,
\[
\int dzg(z)=\int d^2yf(y)=1\;.
\]
To calculate the potential
$\Delta\Phi_c$  of the source (\ref{cyldens}) one
 integrates 
 the density profile $\rho_c(x)$ of the source with the potential
induced by a point-like source,
\begin{equation}
\label{genfinalsize}
\Delta\Phi_c(x)=\int d^3x'\rho_c(x')\Delta\Phi_\delta(x-x')
\end{equation}
Let us begin with the case 
\[
      t/t_c\gg 1 \;, \;\;\; \mbox{i.e.} \; T \gg 1
\]
 and consider the track region, where
the potential is exponentially large. Making use of Eq.~(\ref{largepoint2}) 
one gets
\be
\label{thicktrack1}
\Delta\Phi_c=-\sqrt{\pi}{G\mu\over 2vt_cL^2H}\e^{T/2}A(y)B(z)
\ee
where the functions $A$ and $B$ are given by the following integrals
\be
\label{Aintegral}
A(y)=\int_{|y-y'|\lesssim r_cT}d^2yf(y'/L)J_0\l{|y-y'|\over \sqrt{2}r_c}\r
\ee
and
\be
\label{Bintegral}
B(z)=\int_{z-vt}^z{dz'\over \sqrt{T(t,z-z')}}g(z'/H)\e^{z'/2vt_c}
\ee
In what follows we adopt a natural assumption that $L\sim H$.

Let us first consider the case of large sources,
\be
\label{largesource}
L,H\gg r_cT
\ee
The condition (\ref{largesource}) implies that {\it the diameter of the
track of the source is equal to the size of the source itself}.

In the integral (\ref{Aintegral}), one can substitue 
$f(y'/L)$ by
$f(y/L)$. As a result one obtains  in this case
\be
\label{largA}
A(y)\simeq 2\pi f(y/L)\int_0^{Tr_c}dy'y'J_0\l{y'\over \sqrt{2}r_c}\r={2\sqrt{2}
\pi r_c^2 T} 
 f(y/L)J_1\l{T\over\sqrt{2}}\r
\ee
To estimate the function $B(z)$ let us take the source with the step profile
in $z$-direction,
\[
g(z/H)=\left\{
\begin{array}{c}
1\;,\;\; 0<z<H\\
0\;,\;\; z<0\mbox{ or }z>H
\end{array}
\right.
\]
Then
\be
\label{largB}
B(z)\simeq {b_0\over \sqrt{T(t,\tilde{z})}}\e^{\tilde{z}/ 2vt_c}\;,
\ee
where
\[
\tilde{z}=\mbox{min}(H,z)\;,\;\;b_0=\mbox{min}(H,z,vt_c)
\]
Plugging Eqs.~(\ref{largA}), (\ref{largB}) into the general 
expression (\ref{thicktrack1}) one finally obtains the following result
for the potential inside the track of the large homogeneous source
\be
\label{largefinal}
\Delta\Phi_c\simeq -(2\pi)^{3/2}{G\mu r_c^2T(t,z)b_0 f(y/L)\over L^2Hvt_c
  \sqrt{T(t,z-\tilde{z}) }}\exp{T(t,\tilde{z})\over 2}J_1\l{T(t,z)
  \over\sqrt{2}}\r 
\ee 
By comparing  Eq.~(\ref{largefinal}) to
Eq.~(\ref{largepoint2}) one observes that the 
potential inside the track of the large
source is suppressed as compared to the potential inside the track of a
point-like source of the same mass by a factor 
\be
\label{supp1}
{\Delta\Phi_c\over \Delta\Phi_\delta}\simeq  {r_c^2T^{1/2}b_0\over L^2
  H}<{r_c^2T^{1/2}\over L^2}
\ee
The meaning of this result is easy to understand qualitatively. 
At a given point inside the track, only a part of the source, whose 
transverse size is of order
$(r_cT)$ contributes to the potential. This yeilds  a geometrical 
suppression factor $(r_cT/L)^2$. The suppression 
by an extra factor of $T^{-3/2}$ comes
  from the integration of the rapidly oscillating function in
  Eq.~(\ref{largA}). 

We see that for large homogeneous sources the transverse profile of the
potential inside  the track 
is just proportional to the transverse profile of the source, i.e. this
potential is slowly changing  in the transverse direction. 
This potential slowly oscillates with the
exponentially growing amplitude along the trajectory of the source.

Let us discuss now what happens with  sources of  smaller size. Clearly,
if the size of the source is  much smaller 
than $r_c$,  then the point-like
approximation discussed above is applicable. So, here
we consider sources of the intermediate size,
\[
r_c\ll L,\;H\ll r_cT
\]
In this case it is no longer  possible to pick the profile of the source 
$f(y'/L)$ out of the integral in Eq.~(\ref{Aintegral}). To estimate the
suppression factor,  let us take this profile in the form of the
step function. Also, for simplicity, let us calculate just the value of
the function $A(y)$ at the origin $y=0$,
\be
\label{origin}
A(0)=\int_0^Ld^2y'J_0\l{y'\over \sqrt{2}r_c}\r=2\sqrt{2}\pi
Lr_cJ_1\l{L\over\sqrt{2}r_c}\r 
\ee
For the function $B(z)$ one can still use the
estimate (\ref{largB}). Then, proceeding
as above, we obtain for the suppression factor in this case
\be
\label{supp2}
{\Delta\Phi_c\over \Delta\Phi_\delta}<\l{r_c\over L}\r^{3/2}
\ee

Let us discuss now  the effect of the finite size of the source in the
wave zone. It is rather clear that suppression similar to 
that discussed above
should be present in this region as well.
To see this, let us plug the 
expression (\ref{weaklimit}) for the potential
of the point-like source into Eq.~(\ref{genfinalsize}).  One finds that the
potential of the cylindrical source in the wave zone is given by
\be
\label{cylwz}
\Delta\Phi_c=-{G\mu r_c^2\over vt_cL^2H}A(y,z)\;,
\ee
where
\be
\label{Ayz}
A(y,z)=\int {d^2y'dz'f(y'/L)g(z'/H)\over |y-y'|^2}
T(t,z')\sin{ |y-y'|^2\over 4r_c^2T(t,z')}
%
\ee
Now, for simplicity, we restrict our consideration to the case when the
distance from the source trajectory to the observer is much larger than the
size of the source, $y\gg L,H$. Then, in the non-oscillating part of
the integrand in Eq.~(\ref{Ayz}) one can set $|y-y'|\sim y$. In the resulting
integral one performs the integration over the polar angle in the
$y'$-plane, and obtains
\be
\label{Ayz1}
A(y,z)={2\pi\over y^2}\int y'dy'dz'f(y'/L)g(z'/H)
T(t,z-z')J_0\l{2yy'\over 4r_c^2T(t,z-z')}\r\sin{ |y-y'|^2\over 4r_c^2T(t,z-z')}
\ee
Now, to estimate
the value of the integral over the radial variable $y'$
let us take the profile in the transverse direction to be a Gaussian,
\[
f(y/L)={1\over\pi}\e^{-y^2/L^2}\;.
\]
Then the integral over $y'$ gives two different terms. One of them comes from
the integration over large values of $y'$ and is exponetially suppressed at
large distances from the source, the second comes from the integration over
a tiny region near the origin, $y'\lesssim {r_c^2T\over y}$. Schematically,
barring coefficients of order one and constant phases of oscillations, one has
\be
\label{Ayz2}
A(y,z)={2\pi\over y^2}\int dz'g(z'/H) \left[
\l r_c^2T'\over y\r^2\sin{y^2\over 4r_c^2T'}
+T'^{3/2}r_c^2\e^{-y^2/L^2}\sin{{4y^2r_c^2T'\over L^4}}\right]
\ee
where $T'$ stands for $T(t,z-z')$.

Using Eq.~(\ref{Ayz2}) one arrives at the following upper bound on the
absolute value of the function $A(y,z)$,
\be
\label{absestimate}
|A(y,z)|\lesssim {2\pi H\over y^2} \left[
\l r_c^2T\over y\r^2
+T^{3/2}r_c^2\e^{-y^2/L^2}\right]
\ee
Plugging this upper bound into Eq.~(\ref{cylwz}) and comparing the result with 
the potential (\ref{weaklimit}) for the point source, one finds that in the
wave zone
\be
\label{wavebound}
{\Delta\Phi_c\over \Delta\Phi_\delta}<\l r_c^2T\over Ly\r^2+T^{1/2}{r_c^2\over
  L^2}\e^{-y^2/L^2}
\ee
i.e.,  the suppression of the potential in the wave zone 
relative to the potential of the point
  source of the same mass is even stronger than in the track region.
\section{Discussion and conclusions}
\label{evidence}
Now we are at a point to discuss possible observational
signatures of  ghost condensate, taking into account the effect of non-zero
velocity of all stellar objects in the rest frame of ghost condensate. Let
us start with a few preliminary remarks. From the phenomenological point of
view, ghost condensate  (at the linearized level) has 
the characteristic time and length scales $t_c$ and $r_c$. These parameters
are related to the two microscopic parameters,  mass scale $M$ and
dimensionless parameter $\alpha$, as written in Eqs.~(\ref{rc}), (\ref{tc}).
The allowed deviation of the latter parameter from unity is determined by the
amount of fine-tuning that one  is ready to introduce in the theory.
We are going to be rather generous in this respect; in fact, our
 discussion is quite flexible 
 and as large values of $\alpha$ as $10^{10}$
 will not affect it significantly. 

An exhaustive phenomenological analysis of the theory would have resulted
 in the
exclusion plot in the $(t_c,r_c)$-parameter space, and in the detailed
discussion of the  characteristic observational signatures for different
allowed regions. 
We believe that such an analysis deserves a separate publication.
Our purpose here is to discuss 
 qualitative features of the ghost condensate phenomenology, 
stressing the crucial role of the
 effect of  
finite velocity. 
Our  claims are the following.
\begin{enumerate}
\item It is very unlikely to observe ghost condensate with
$t_c\sim t_U$,
where $t_U$ is the present age
of the Universe. In other words, it is crucial
for the observability of  ghost condensate that it enters 
the regime in which tracks with
exponentially enhanced field are present.
\item The tracks of compact massive objects
  (stars) become 
pronounced earlier than the tracks 
of the supermassive objects of small density (galaxies).
\item The chance to observe ghost condensate is larger for larger
  values of $r_c$.
\item Relatively promising ways of searching 
for tracks in ghost condensate are: (i) search 
for ``mad'' stars which feel the gravitational field of
  the tracks of other stars; (ii)
microlensing
  observations and (iii) gravitational wave
  experiments.
\item It may happen that $t_c$ is so small 
that tracks of some objects are already in the
  non-linear (quantum?) regime and, still, we have 
not noticed the presence of
   ghost condensate so far. 
Consequently, it may happen that to fully understand
  the phenomenology of ghost condensate one needs the details
of the UV completion in the ghost sector.
\end{enumerate}

Let us first explain why it is unlikely to observe  ghost condensate for
large characteristic time scales, $t_c\gtrsim t_U$. In this regime,
tracks with exponentially enhanced field 
did not have enough time to develop, so the only
potentially
observable effects are due to the gravitational
potential in the wave zone. Let us
first assume that the characteristic size $r_c$ is somewhat larger than the
size of a typical star like the Sun,
\[
r_c\gtrsim 10^6\;\mbox{km}\;
\]
This actually implies that the
parameter $\alpha$ is quite large, $\alpha\gtrsim
10^5$; however, as we will see, the chance to detect ghost condensate
is even lower for
smaller values of $r_c$. 
Then, to estimate the ``extra'' gravitational potential  $\Delta\Phi$
(see, Eq.~(\ref{NDel})) of a star 
whose trajectory was at a distance $y_0$ to the current
location of the Earth (in the rest frame of ghost condensate)
 we make use of the expression (\ref{weaklimit}) for
the potential of a point-like source in the wave zone,
\be
\label{weakstar}
\Delta\Phi\sim 10^{-20}
\l \mu\over M_{\odot}\r
\l{ 10^{-3}\over v}\r
\l{r_c\over    y_0}\r^2
\sin{y_0\Delta y\over 2r_c^2}
\ee
where we set $T=1$, expanded the phase of oscillations in Eq.~(\ref{weaklimit})
near a given space-time point setting
\[ 
y=y_0+\Delta y
\] 
neglected  extremely 
slow variation of this phase in time and in $z$-direction and  dropped  the
constant shift of this phase. Equation (\ref{weakstar}) applies to the
rest frame of ghost condensate, while in the rest frame of the Earth
the gravitational field has the form of a wave
of the amplitude
$\sim 10^{-20}$  and frequency 
\be
\label{nu}
\nu= {v_r y_0\over 4\pi r_c^2}\simeq 2\cdot 10^{-5}\; \mbox{Hz}\l
{v_r\over
  10^{-3}}\r \l{y_0\over r_c}\r\l{10^6\;\mbox{km}\over r_c}\r
\ee
where $v_r$ is the Earth velocity in the direction transverse 
to the star trajectory. Note that this is a scalar wave unlike the tensor waves
of the Einstein theory.
The gravity waves of such a 
low frequency are in principle accessible to the LISA project (see, e.g.,
Refs.~\cite{Thorne:1995xs,Lobo:2002pr} for  reviews of the gravitational
wave experiments), however, its sensitivity at these frequences is at the
level\footnote{It is worth noting that gravitational
wave experiments are sensitive not to the amplitude $\Delta\Phi$ of the
gravitational wave itself, but to the product $\sqrt{n}\Delta\Phi$, where $n$
is a number of cycles produced in a logarithmic band about a given
frequency. In the case at hand $n\sim\nu y_0/v_r=y_0^2/(4\pi r_c^2)$. 
This comment is practically irrelevant for our discussion.}
 $\Delta\Phi\sim 10^{-17}$.

Furthemore, it is straightforward to see that the probability $p(r_c)$
to have even this weak 
signal is extremely low. 
Indeed, to estimate this probability, note
first that the probability ${\cal P}(r_c)$
that  the distance from the current position of the Earth to the nearest
 trajectory of a star is smaller than $r_c$, is given by
\be
\label{probest}
{\cal P}(r_c)\sim {N_{st}N_gvt_Ur_c^2\over r_U^3}\sim 10^{-15}
\ee
where $N_{st}\sim N_g\sim 10^{11}$ are, respectively,
 the number of stars in a typical galaxy and
the number of galaxies in the Hubble volume; $r_U\sim 10^{28}$ cm, $v\sim
10^{-3} $ and $r_c\sim 10^6$~km. For long enough period of
observation
$t_o$, such
that the Earth travels a distance $L_E$ much larger than $r_c$, the probability
$p(r_c)$ is somewhat larger,
\[
p(r_c)\sim {\cal P}(r_c){L_E\over r_c}\sim 10^4\cdot
{\cal P}(r_c)\l {t_o
\over 1\;\mbox{yr} }\r\l{v\over 10^{-3}}\r\l{10^6\;\mbox{km}\over r_c}\r
\]
Since the amplitude in Eq.~(\ref{weakstar}) rapidly decreases at $y_0
> r_c$, this expression determines the probability of having a 
non-negligible signal. Clearly, this probability is quite low.

The sensitivity of  gravitational wave detectors 
 is higher in the higher frequency bands, so one 
might expect better signal for smaller $r_c$, and hence higher
frequency
$\nu$.
For instance, LISA will have  sensitivity at the level
$\Delta\Phi\sim 10^{-20}$ in the frequency range $10^{-3}\div 10^{-1}$~Hz.
The waves of
amplitudes and frequencies in this range would be generated
for 
$y\sim r_c\sim 1000$~km. In this case,
in order to avoid the suppression of the
amplitude due to the effect of the finite size of the source
(section \ref{thicksub}), one should consider very compact sources like neutron
stars. But the above estimate shows that we should be extremely
lucky to have a
trajectory of a neutron star at a distance of order 1000 km from the LISA
facility.

Let us now discuss the effects  from  objects of larger
size, e.g. galaxies. 
One could expect two different types of signatures on
these large distance scales. The first is 
the gravitational wave signals like those
discussed above. The second is the
modification of the gravitational dynamics at large
scales due to extra contributions to the Newtonian potential.
However, it is
straightforward to see that the effect of averaging discussed in section
\ref{thicksub} kills all these signatures unless the value of the
characteristic length scale $r_c$ (and, correspondingly, of the parameter
$\alpha$) is extremely large. Indeed, due to the large mass of a galaxy
there is an extra factor of order $10^{12}$, as compared to a star,
 in the estimate (\ref{weakstar}) 
for the gravitational potential. However, Eq.~(\ref{wavebound}) tells us that 
there is an extra suppression by at least a factor 
\be
\label{halofield}
{r_c^2\over L_g^2}\sim 10^{-30}\l{r_c\over
  1000\;\mbox{km}}\r^2\l{30\;\mbox{kpc} \over L_g}\r^2
\ee
where $L_g$ is the size of a galaxy. We see that the effects of ghost
condensate are indeed very small for large objects of small density.

The above arguments show that  chance to detect  
ghost condensate is very low if the characteristic time scale
$t_c$ is longer than, or equal to
the present age of  the Universe. This forces us to discuss
the regime $t_c\lesssim t_U$.
This
regime is rather dangerous, as the gravitational and ghost fields
grow exponentially  inside the tracks.  
Still, let us make a few general
remarks,  postponing the detailed discussion of the ghost condensate
phenomenology in this regime for future. 

 If the 
 ratio $t_U/t_c$ is large enough, the tracks of  galactic halos
are very pronounced (say, the gravitational potential in the track
is comparable to the typical gravitational potential between two interacting
galaxies).
The
fraction of the Universe filled by these tracks is  estimated as
\be
\label{halofraction}
{\Delta V_h\over V_U}\sim N_g{r^2_{halo}vt_u\over t_u^3}\sim 0.05 \l{N_g\over
  10^{11}}\r \l{r_{halo}\over 100\mbox{ kpc}}\r^2\l{v\over 10^{-3}}\r
\ee
where we estimated the size of a halo of a typical galaxy as 100 kpc. It is
  unlikely that
  this effect would have been  unnoticed. For instance, the dynamics of 
a sizeable number of galaxies would have been
affected by these tracks. 

For not so large $t_U/t_c$ the situation is more interesting.
There is a range of parameters in which tracks of galaxies are
unnoticeable, but star tracks are strong.
As an example, one finds
from Eqs.~(\ref{weakstar}) and (\ref{halofield}) (assuming $r_c=1000$~km
for definiteness) 
that for
\[
{t_U\over t_c}\sim 2 \ln{10^{20}}\sim 92
\]
the gravitational potentials in the tracks of  neutron stars are of
order one, while the gravitational potentials in the tracks 
of  typical galactic halos are still of order $10^{-18}$ which is more than
10 orders of magnitude smaller than the gravitational potential of a typical
galaxy at a distance of order 1 Mpc. Consequently, there is an interesting
range of the characteristic time scales,
\be
\label{interesting}
0.01 \lesssim {t_c\over t_U} <1
\ee
in which star tracks, but not galactic tracks, are
pronounced (up to $\Delta\Phi\sim 1$). 

Normally, the existence of exponentially growing 
field would mean that an
extreme fine-tuning of the parameters is needed to have potentially observable
effects without ruling out the theory completely.
Clearly, to have $t_c$ in the range (\ref{interesting}) one needs some
fine-tuning, but not that strong as one might expect. 

We suggest here three potential signatures of the star tracks.
 First, 
one can search for  ``mad'' stars, intersecting the tracks of other stars,
so that their motion is strongly affected by the gravitational field of
the track (probably, just for a short period of time). Second, 
it seems possible to observe tracks in the
microlensing experiments (see, e.g., 
Ref.~\cite{Roulet:1996ur} for a review of microlensing experiments) due to the
variation of the visible luminosity of a background star when the line
of sight  crosses the track. Finally,  
the gravitational wave detetors may detect a signal from the track, as
discussed above.

To observe  mad stars one needs a galaxy whose disc is intersecting a track
of the disc of another galaxy at the moment of observation. Plugging the
typical disc size $r_{disc}\sim 10$~kpc instead of $r_{halo}$ into
Eq.~(\ref{halofraction}) we find that this happens for approximately 
 one galaxy of a thousand. To get a feeling of numbers let us also estimate
the fraction of the volume of such a galaxy filled by the
star tracks,
\be
\label{starfraction}
{\Delta V_{s}\over V_{galaxy}}\sim N_s\l {r_s\over r_{disc}}\r^2\sim
10^{-12}\;,
\ee
where $r_s\sim 10^6$~km is the radius of a typical star. This estimate is not
very optimistic as it implies that there is about one mad star in the
galaxy at each moment of time. Note however, that one can significantly
enhance the success probability by performing  monitoring of stars for a long
period of time. Also it would be interesting to check whether 
 trapping of  a star by the gravitational field of a track is possible.
Finally, the estimate (\ref{starfraction})  assumes 
 that the parameter $r_c$ is smaller than
$r_s$ so that the diameter of the star track is equal to the size of the
star. At larger values of $r_c$ this fraction is enhanced by a factor of
$(r_c/ r_s)^2$.

Similar considerations 
 show that chance to observe two other signatures 
(microlensing and gravity waves) 
 are rather low  for
 $r_c<r_s$.

The above estimates show that it may happen, that while we have not
noticed the presence of ghost condensate yet, the gravitational and ghost
fields in the tracks of some dense objects are already very strong. Note that
it follows from Eq.~(\ref{action}) that
gravitational field $\Delta\Phi\sim 1$ corresponds to  ghost field
$\pi_c\sim M$, so  it is unclear whether  non-linear classical dynamics will
result in some stable non-linear classical track solution or  strong
quantum effects will start operating at this point. [In particular, there is a
danger that the points of the field space where excitations of  ghost
condensate are ghosts (i.e., have wrong sign in front of 
kinetic term) become
accessible inside the tracks.] 
It seems difficult to deduce what is happening in
this regime without knowing the UV completed theory of  ghost condensate
(in other words, without constructing a complete 
analog of the Higgs mechanism for
gravity and not just a non-linear sigma-model).


To conclude, let us mention some other open
questions and further directions of research. 

First, in our treatment we
did not take into account the expansion of the Universe. 
In the interesting case (\ref{interesting})
this assumption seems 
justified because at the initial, rather short cosmological
stage when the Hubble rate was higher
than the rate of the development of instabilities in  ghost condensate
$t_c^{-1}$, the Hubble friction prevented the instabilities to grow.
Also, the Universe was essentially homogeneous during the fast period of
expansion, while the tracks discussed here emerge due to the inhomogeneities.
Still, we believe that this question deserves further study. Especially
interesting would be to check the possibility to have sizeable tracks of 
inhomogeneities which could be present before inflationary stage.

Second, our discussion was purely non-relativistic.  However, one could argue
that there is a possibility that the whole observed part of the Universe is
moving with respect to the rest frame of ghost condensate with  velocity
close to the speed of light. 
It would be interesting to check whether this is a consistent
possibility,
or 
one of the
effects of the Hubble friction is to slow down the velocity with respect to
the rest frame of  ghost condensate, so the non-relativistic approximation
is always justified. 
In the latter case it would be
natural to assume that the
rest frame of  ghost condensate coincides with the rest frame of the CMB.
If true, this assumption would imply 
that the motion with respect to  ghost
condensate is entirely determined by the peculiar velocities of galaxies and
galactic clusters. Then it would be interesting 
to reconstruct the actual map of
galactic tracks at least in the local part of the Universe. Such a map 
would
significantly decrease the uncertainties in the observational predictions of
the model.

It is worth  noting also,  
that the  effect of finite velocity of an observer with respect to the
rest frame of ghost condensate should also be taken into account while
discussing the limits on the direct coupling between matter fields and ghost
condensate,  e.g., coming from the spin-dependent 
forces~\cite{Arkani-Hamed:2003uy} mediated by
 ghost condensate.

Finally, 
ghost condensate may be considered as a particular case of the
$k$-essence models~\cite{Armendariz-Picon:2000dh,Armendariz-Picon:2000ah}. 
It is interesting to check whether the effects
similar to those
discussed here can be present and pronounced in the
models of $k$-essence with cosmologically relevant parameters.

\section*{Acknowlegements}
We thank R.~Rattazzi and I.~Tkachev for stimulating discussions 
and V.~Rubakov for useful correspondence and critical reading 
of the manuscript.
\section*{Appendix A: Asymptotics of the ``propagator'' $I(T,R)$}
\label{ITRap}
\begin{figure}[t]
\begin{center}
\epsfig{file=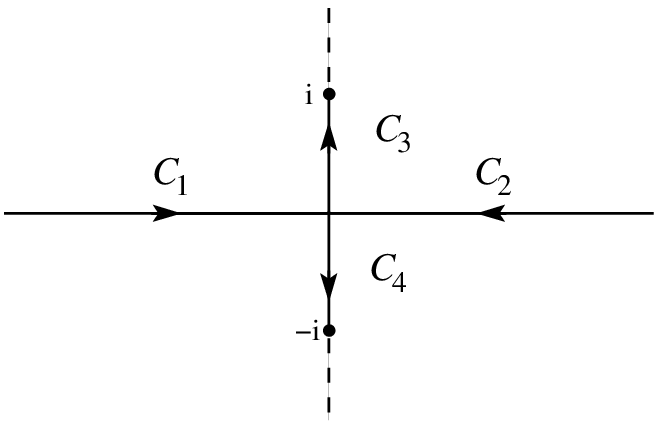,height=3cm,width=6cm}
~~~~~~~~~~~
\epsfig{file=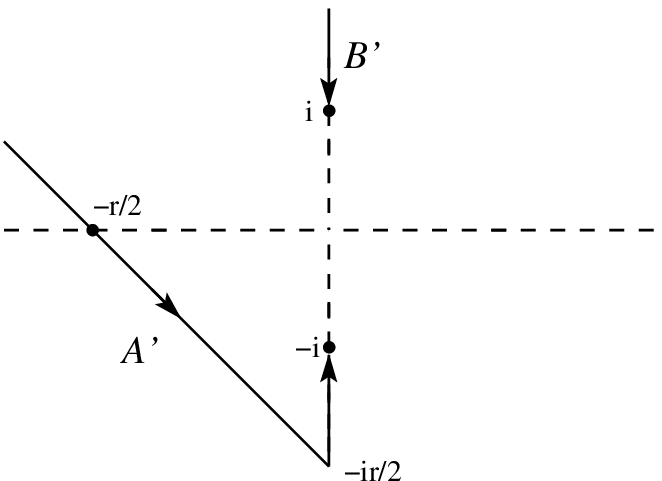,,height=3cm,width=6cm}
\caption{}
\label{contours}
\end{center}
\end{figure}

The purpose of this Appendix is to find the asymptotics of the propagator
$I(T,R)$ in physically interesting regions.  For this purpose, let us first
introduce a new integration variable
\[
v=\sqrt{u^2-1}
\]
Then it is straightforward to check that the function $I(T,R)$ takes  form
of the following contour integral
\begin{equation}
\label{vintegration}
I(T,R)={1\over 2}\mbox{Re}\int_{\cal C}
{dv\over\sqrt{1+v^2}}\e^{i\sqrt{1+v^2}(Tv+R)}\;,
\end{equation}
where the integration contour ${\cal C}$ consists of four segments (see
Fig. \ref{contours}a)),

\[
{\cal C}={\cal C}_1+{\cal C}_2+{\cal C}_3+{\cal C}_4
\]
with
\[
{\cal C}_1=(-\infty, 0)\;,\;\;{\cal C}_2=( \infty,0)\;,\;\;{\cal C}_3=( 0,i
)\;,\;\;{\cal C}_4=( 0,-i)
\]
Now, to perform the  saddle point integration we need the extrema of
the exponent
\[
f(v)=i\sqrt{1+v^2}(Tv+R)
\]
of the integrand in Eq.~(\ref{vintegration}).
This function has two stationary points, given by
\[
v_\pm={-r\over 4}\pm\sqrt{r^2-8}\;,
\]
where we introduced the
 ratio 
\[
r\equiv {R\over T}
\]
  At this point it is
convenient to consider separately two limiting cases $r\ll 1$ and $r\gg 1$.
\subsection*{A1: $r\ll 1$}
 Let us start with the case
$r\ll 1$. Here, both saddle points are near the imaginary axis,
\[
v_\pm\approx{-r\over 4}\pm {i\over \sqrt{2}}\;,
\]
Now, at $T\gg 1$ the function $f$ takes large positive value at $v_-$.  
Consequently, in this range of parameters one finds
that the ${\cal C}_4$-part of the contour ${\cal C}$ gives exponentially large
contribution to $I(T,R)$,
while other parts of the contour give at most contributions of order one.
Leaving only the exponentially large piece we obtain the following saddle
point approximation for the function $I(T,R)$,
\begin{equation}
\label{largeTsmallR}
I(T,R)={1\over 2}\sqrt{\pi\over T}\exp\l{T\over 2}-
{R^2\over 8T}\r\sin{R\over\sqrt{2}}\;,\;\; R\ll T\;,\;\; T\gg 1
\end{equation}
This exponetially large term agrees with that found in 
Ref.~\cite{Arkani-Hamed:2003uy}.

In the opposite case $T\ll 1$ function $I(T,R)$ tends 
to zero as $const\cdot (T
R)$ as is obvious from the original expression, Eq.~(\ref{ITR}).  
\subsection*{A2: $r\gg 1$}
Let us now consider the opposite limit $r\gg 1$ (more precisely, $R/T\gg
\sqrt{8}$). Here both saddle points $v_{\pm}$ are on the real axis, 
\[
v_+\approx -{1\over r}\;,\;\; v_-\approx -{r\over 2}\;.
\]
First, let us note, that for $R\ll 1$ the
function $I(T,R)$ again tends to zero as
$const\cdot (T R)$. 

Let us now turn to the regime $R\gg 1$.
It is convenient to combine contours ${\cal C}_1$, ${\cal C}_4$ into a
single contour ${\cal A}$ and contours ${\cal C}_2$, ${\cal C}_3$ into 
a single
contour ${\cal B}$ and deform contours  ${\cal A}$ and ${\cal B}$
to contours  ${\cal A}'$ and ${\cal B}'$ as shown in Fig.~3. Then one can
check  that  
contributions to $I(T,R)$ that do not have exponential suppression come from 
the part of the contour  ${\cal A}'$ in the vicinity of the saddle point 
$v_-$ and from the parts of the contours ${\cal A}'$ and ${\cal B}'$
in the vicinity of points $\mp i$. Furthemore, it 
is straightforward to check that the 
two  latter contributions actually cancel out,  so one is left  with the
saddle point   contribution which is now given by
\begin{equation}
\label{largeR}
I=\sqrt{\pi T\over 2}
{\cos\l
{R^2\over 4T}+{T\over 2}\r
+
\sin\l{R^2\over 4T}+{T\over 2}\r\over R}\;,\;\;R\gg\sqrt{8}T\;,\;\;
R\gg 1
\end{equation}
To summarize, Eqs. (\ref{largeTsmallR}), (\ref{largeR})
provide a good
approximation for the function $I(T,R)$ in all physically interesting regions.

\end{document}